\documentclass[openacc,capitalize]{rstransa}

\pdfoutput=1

\jname{rsta}
\Journal{Phil. Trans. R. Soc}

\usepackage[utf8]{inputenc}
\usepackage[T1]{fontenc}
\usepackage[english]{babel}

\usepackage[active]{srcltx}
\usepackage{bm,microtype}
\usepackage[usenames,dvipsnames]{xcolor}
\usepackage{mathrsfs,mathdots}
\usepackage[centercolon=true]{mathtools}
\usepackage{braket,enumitem}
\usepackage{suffix,xifthen,xspace}
\usepackage{booktabs,bookmark}

\usepackage{csquotes}
\usepackage[capitalize]{cleveref}
\usepackage[compress,numbers]{natbib}

\newcommand*{\swap}[2]{\let\temp#1 \let#1#2 \let#2\temp \let\temp\relax}

\swap{\epsilon}{\varepsilon}
\swap{\theta}{\vartheta}
\swap{\phi}{\varphi}
\DeclarePairedDelimiter{\abs}{\lvert}{\rvert}

\DeclarePairedDelimiter{\card}{\lvert}{\rvert}
\DeclarePairedDelimiter{\len}{\lvert}{\rvert}

\DeclarePairedDelimiter{\paren}{\lparen}{\rparen}
\DeclarePairedDelimiter{\sparen}{[}{]}

\DeclarePairedDelimiter{\Gen}{\langle}{\rangle}
\def\pp#1{\paren*{#1}}

\newcommand*{\numberset}{\mathbb}
\newcommand*{\N}{\numberset{N}}
\newcommand*{\Z}{\numberset{Z}}

\newcommand*{\C}{\numberset{C}}

\newcommand*{\im}{i}

\renewcommand*{\vec}[1]{{\bm{\mathrm{#1}}}}

\def\v#1{\vec{#1}}
\def\bh{\v h}
\def\bx{\v x}

\newcommand*{\Hilbert}[1][H]{\mathcal{#1}}
\def\H{\Hilbert}
\newcommand*{\Aut}[1]{\mathrm{Aut}\paren{#1}}
\WithSuffix\newcommand\Aut*[1]{\mathrm{Aut}\paren*{#1}}

\def\sp#1{\mkern#1mu}
\def\deq{\coloneqq}
\newcommand{\CM}[1][s]{\mathcal{M}_{#1}(\C)}
\def\CE{\mathcal{E}}
\def\Neigh{\mathcal N}
\def\CC{\mathcal{C}}
\def\CA{\mathcal{A}}

\def\FB{\mathfrak{B}}
\def\FS{\mathfrak{S}}
\def\FW{\mathfrak{W}}
\def\FT{\mathfrak{T}}
\def\FZ{\mathfrak{Z}}
\newcommand*{\PC}[1][x]{\CC_t(#1)}
\newcommand*{\paths}[1][x',x]{\Lambda_t(#1)}

\def\ws#1{w^{(#1)}}
\def\wn{\left(\ws1,\dots,\ws{n}\right)}

\def\wtre{\left(\ws1,\ws2,\ws3\right)}
\def\Bin#1{\FB_{#1}} % Binary strings of a given length.
\def\BS#1#2{\FS_{#1}(#2)} % Binary strings with fixed number of ones.
\def\BT#1#2#3{\FS_{#1}^{#2}(#3)} % Binary tuples with fixed number of ones each.
\def\BW#1#2#3{\FW(#1,#2,#3)} % Binary strings partitioning.
\def\BWn{\BW KHn} % Binary strings partitioning default.
\def\BST#1#2#3#4{\FT_{#1}(#2,#3,#4)}
\def\BSTn{\BST{aa'}tKn}
\def\BZ#1#2#3{\FZ_{#1}^{(#2)}(#3)}
\def\BZn#1{\BZ{b}{#1}{J,k}}

\newcommand*{\tp}[1]{_{\textup{#1}}}

\newcommand*{\mvec}[1]{\begin{pmatrix}#1\end{pmatrix}}

\newcommand*{\Kleene}[1]{#1^\star}
\DeclarePairedDelimiter{\Pres}{\langle}{\rangle}
\newcommand{\Cay}[1]{\Gamma(#1)}

\theoremstyle{definition}
\newtheorem{definition}{Definition}[section]

\theoremstyle{plain}

\newtheorem{proposition}{Proposition}[section]
\newtheorem{lemma}{Lemma}[section]
\newtheorem{corollary}{Corollary}[section]

\newcommand*{\DefaultAcronymStyle}[1]{#1}
\newcommand*{\Acronym}[2][\DefaultAcronymStyle]{\expandafter\def\csname #2\endcsname{#1{#2}\xspace}}
\WithSuffix\newcommand\Acronym*[3][\DefaultAcronymStyle]{\expandafter\def\csname #2\endcsname{#1{#3}\xspace}}

\Acronym{QCA}
\Acronym{QW}
\Acronym{RW}
\Acronym{BCC}

\Acronym*[\emph]{ie}{i.e.}
\Acronym*[\emph]{eg}{e.g.}

\begin{document}

\title{Path-sum solution of the Weyl Quantum Walk in 3+1 dimensions}

\author{G.~M.~D'Ariano$^{1}$, N.~Mosco$^{1}$, P.~Perinotti$^{1}$, A.~Tosini$^{1}$}

\address{%
$^{1}$QUIT group, Dipartimento di Fisica, via Bassi 6, 27100 Pavia, Italy}

\subject{Quantum computing, Particle physics, Quantum physics}

\keywords{Quantum Walks, Path-sum, Path-integral, Quantum Field Theory, Discrete Time}

\corres{Insert corresponding author name\\
\email{paolo.perinotti@unipv.it}}

\begin{abstract}
  We consider the Weyl quantum walk in $3+1$ dimensions, that is a discrete-time
  walk describing a particle with two internal degrees of freedom moving on a
  Cayley graph of the group $\Z^3$, that in an appropriate regime evolves
  according to Weyl's equation. The Weyl quantum walk was recently derived
  as the unique unitary evolution on a Cayley graph of $\Z^3$ that is
  homogeneous and isotropic. The general solution of the quantum walk evolution
  is provided here in the position representation, by the analytical expression of
  the propagator, i.e.~transition amplitude from a node of the graph to another node in a
  finite number of steps. The quantum nature of the walk manifests itself in the interference of
  the paths on the graph joining the given nodes. The solution is based on the binary
  encoding of the admissible paths on the graph and on the semigroup structure of the
  walk transition matrices.
\end{abstract}
%%%%%%%%%%%%%%%%%%%%%%%%%%%

% \begin{fmtext}
% \end{fmtext}

%%%%%%%%%%%%%%% End of first page %%%%%%%%%%%%%%%%%%%%%

\maketitle

\section{Introduction}

Quantum walks (QWs) have been originally introduced as the quantum analog of
classical random walks or Markov chains~\cite{Accardi:1989aa}. A discrete-time
QW~\cite{Grossing:1988aa,Aharonov:1993aa,Aharonov:2001aa} describes a quantum
particle jumping on a graph in discrete time steps, the ``direction'' of the
jumping being conditioned on the internal state of the particle. The latter is
represented in a finite dimensional Hilbert space---the \emph{coin}
space---while  the Hilbert space associated with the positions on the lattice is
generally infinite dimensional.  The total Hilbert space of the walk is the
tensor product of the above ones.
 
Since their first appearance, QWs received an increasing attention in the
literature,  where their main application is the design of quantum
algorithms~\cite{Ambainis:2007aa,Magniez:2007aa,Farhi:2007aa,Childs:2003aa}.  As
an example, efficient search
algorithms~\cite{Magniez:2007aa,Santos:2015aa,Wong:2015aa}  were devised
exploiting the fact that the spread of a localized initial state after $t$ steps
is proportional  to $t$, whereas for the classical random walk the spread is
proportional to  $\sqrt{t}$. Remarkably, any quantum circuit can be implemented
as a QW on a graph, proving that QWs can be used for universal quantum
computation~\cite{Childs:2009aa}.

Recently, in Ref.~\cite{DAriano:2014ae} the authors use QWs in a derivation of quantum field
theory from principles, from which it follows that the graph of the QW is the {\em Cayley graph}  of
a finitely presented group $G$. The QWs reproducing free quantum field theory in Euclidean space
correspond to group $G= \Z^d$ that is Abelian, which is the most common case in the literature.
In particular the simplest QW descending from the principles is the Weyl QW \cite{DAriano:2014ae},
reproducing the Weyl quantum field theory. The QW approach to quantum field theory has been
successfully used in
Refs.~\cite{Arrighi:2013ab,Bisio:2013ab,Arrighi:2014aa,DAriano:2014ae,Bisio:2014aa,Bisio:2015aa,Bibeau-Delisle:2015aa,Bisio:2015ab,Bisio:2015ac},
where non-interacting field theories are studied in a simplified picture in terms of Quantum Walks.

In Sec.~\ref{sec:qwc} we review the definition of QWs on Cayley graphs.  In
Sec.~\ref{sec:weyl-qw} we introduce the Weyl QW and solve its path-sum on the
body-centered cubic three dimensional lattice.

\section{Path-sum approach for quantum walks on Cayley graphs}\label{sec:qwc}

In this section we review the notion of a quantum walk, focusing on the
underlying graph structure, that for QWs representing a homogeneous
dynamics is the Cayley graph of some group.

\subsection{Quantum walks on Cayley graphs}

In an abstract way, a group can always be described in terms of a set of
generators. If the generators are viewed as an alphabet, each word over  the
alphabet is associated to an element of the group.  More precisely,
following~\cite{Epstein:1992aa}, an alphabet is any non-empty set $S$ and a
string over an alphabet $S$ is a map $w \colon \{1,2,\dots,n\} \to S$, for some
non-negative integer $n$.  The length of a string $w\colon \{1,2,\dots,n\} \to
S$ is $\len{w} \deq l(w) \deq n$, and for $n = 0$ there is a unique string
denoted $\epsilon$, called the \emph{empty string} for the alphabet $S$.  Let
$\Kleene{S}$ denote the monoid generated from $S$ by the associative operation
of string concatenation, having $\epsilon$ as its unit. Introducing the alphabet
$S^{-1}$ of formal inverses of $S$, we say that a \emph{word over $S$} is a
string over $S \cup S^{-1}$ and a word is called \emph{reduced} if it does not
contain substrings of the form $xx^{-1}$, with $x$ in $S \cup S^{-1}$.  The set
of all reduced words $F(S)$ can be given a group structure where the group
operation is represented by word juxtaposition,  the unit is $\epsilon$, and the
inverse of a word $w$ is the word consisting of the inverses of the symbols of
$w$ juxtaposed in the reverse order. The group $F(S)$ is called the \emph{free group
over the alphabet $S$}.

Any group $G$ has a presentation $\Pres{S|R}$---$S$ being a set of generators
and $R$ any subset of $F(S)$, whose elements are called \emph{relators}\footnote{The relators $r\in
  R$ in the presentation $\Pres{S|R}$ of a group $G$ are words over $S$ that are equivalent
  to $\epsilon$, namely to the identity.} $\epsilon$---so that $G$ is isomorphic to $F(S) / \Gen{R^{F(S)}}$, where $\Gen{R^{F(S)}}$
denotes the normal closure\footnote{The normal closure $\Gen{R^G}$ of a set $R$ in a group $G$ is the subgroup
generated by $gRg^{-1}$ with $g\in G$.} of $R$ in $F(S)$.  Knowing a presentation
$\Pres{S|R}$ of a group, we can give a graphical representation of it, called
Cayley graph,  capturing both its algebraic and geometric properties. We
have the following definition.

\begin{definition}[Cayley Graph]\label{def:cayley}
Let $G$ be a group and let $\Pres{S|R}$ be a presenttion of $G$.  Then the
\emph{Cayley graph of $G$} corresponding to the presentation $\Pres{S|R}$,
denoted $\Cay{G,S}$, is the coloured directed graph $(G, E, \{c_h\}_{h\in S})$
such that the vertex set is $G$, the edge set is $E = \set{(g,gh) \mid g \in G,
\, h \in S}$ and each $h \in S$ has an associated colour $c_h$.
\end{definition}

We are interested in a discrete unitary evolution over such a graph.  Formally we have the
following definition.

\begin{definition}[Quantum walk] \label{def:qw}
    Let $V$ be a countable set, then a (discrete-time) \emph{quantum walk} on $V$ is a local unitary operator $W$ on the Hilbert space $\H=\ell^2(V)\otimes \mathbb C^s$, namely a unitary operator $W$ such that there exists $k\in \N$ and
    \begin{equation*}
    |\Neigh_x|\leq k,
    \end{equation*}
    where the neighborhood $\Neigh_x$ of $x\in V$ is defined as the set
    \begin{align*}
    \Neigh_x\deq\{y\in V\mid\exists\psi,\phi\in\C^s,\bra y\bra\psi W\ket x\ket\phi\neq0\vee\bra x\bra\psi W\ket y\ket\phi\neq0\},
    \end{align*}
    and $\{\ket x\}_{x\in V}$ is the canonical orthonormal basis in $\ell^2(V)$.
\end{definition}

In this context a \QW on the set $V$ can be specified providing the map $\CE\colon V\times V\to\CM$ associating to each pair of sites $(x,y)$ a matrix $\CE(x,y)\deq\bra y W\ket x$, called \emph{transition matrix}, representing the transition amplitude from $x$ to $y$. The unitarity requirement in terms of transition matrices amounts to
   	\begin{equation*}
    	\sum_{z \in V} \CE(z,x)\CE(z,y)^\dagger =
    	\sum_{z \in V} \CE(x,z)^\dagger\CE(y,z) =
    	\delta_{xy} I_s,
   	\end{equation*}
    $I_s$ denoting the identity on $\C^s$.
The locality constraint, on the other hand, becomes
    \begin{equation*}
    	\forall\sp1 x \in V, \quad
    	\card{\set{y \in V | \CE(x,y) \neq 0 \lor \CE(y,x) \neq 0}} \leq k.
    \end{equation*}
    Furthermore, a sequence $\psi\colon \N \to \H$ is a solution of the quantum walk $W$ if it satisfies the
    following update rule for a given initial condition $\psi(0) \in
    \H$:
    \begin{equation}\label{eq:upd-rule}
	    \ket{\psi_x(t+1)}= \sum_{y \in V} \CE(y,x) \ket{\psi_y(t)}, \quad
	    \forall\sp1 x \in V, \; \forall\sp1 t \in \N,
    \end{equation}
    where $\ket{\psi_x(t)}\in\mathbb C^s$ is defined in such a way that $\ket{\psi(t)}=\sum_{x\in V}\ket x\ket{\psi_x(t)}$.
    
A \QW $W$ carries a natural graph structure defined by the non-null transition
matrices, namely the directed graph $(V, E)$ where $V$ is the vertex set and $E$
is the set of edges $E = \set{(x,y) \in V \times V | \CE(x,y) \neq 0}$.  It has
been shown~\cite{DAriano:2014ae} that the assumption of \emph{homogeneity} of
the \QW $W$---\ie the vertices cannot be distinguished by the walk
dynamics---entails that the underlying graph is actually the Cayley graph
$\Cay{G,S}$ of a group $G$.
%In such a case, we further assume that the connection set $S$ is a generating
%set for $G$ and it is symmetric, namely $S = S^{-1}$.
The group $G$ acts on $\ell^2(G)$ by means of the right-regular representation
$g \mapsto T_g$ as $T_g \ket f \deq\ket {fg^{-1}}$.  The homogeneity condition
entails also that the transition matrices are independent of the location:
$\CE(gh,g)=\CE(g'h,g')$, $\forall g,g'\in G$ and $\forall h\in S$, allowing the
choice $\CE(gh,g) = A_h \in \CM$.  Hence the walk operator $W$ can be written as
\begin{equation*}
    W = \sum_{h \in S} T_h \otimes A_h.
\end{equation*}

The scope of this paper is the study of the QW evolution in terms of a path-sum,
recalling the Feynman's formulation of Quantum Mechanics.  Such an approach has
been effective to obtain the exact analytic solution in position space in a
number of cases
\cite{Ambainis:2001aa,Konno:2005aa,DAriano:2014ad,DAriano:2015aa}, giving a
background for possible generalizations of the method on general graphs.  In the
present work we will review the method in the setting of Cayley graphs and we
will present a solution for the Weyl QW.

\subsection{Quantum walk evolution as a path-sum}

Let us consider now a \QW whose transition matrices are given by the map
$\CE\colon V\times V \to \CM$ and whose associated graph is $(V,E)$.  From
\cref{eq:upd-rule} one can readily write the evolution of a given initial
configuration $\ket{\psi(0)}\in\H$ as a path-sum, which can be viewed as a
discrete version of Feynman's
path-integral~\cite{Feynman:1948aa,Feynman:1965aa}.  The iteration of the
one-step update rule~\eqref{eq:upd-rule} leads to an expression of the solution
at time $t$ in terms of the sum over all the paths $\sigma$ of length $t$ with
fixed endpoints $x'$ and $x$, that is $\sigma =(e_1,e_2,\dots,e_t)$, where $e_i\deq (y_{i-1},y_{i})$ with the
the identifications $y_0 \equiv x'$ and $y_t \equiv x$.  We denote the set
of all such paths as $\paths$ and $\PC \deq \set{x' \in G | \paths \neq
\emptyset}$ is the slice at time $t$ of the past causal cone of $x$.  In this
way the equation for the evolved state $\ket{\psi_x(t)}$ takes the form
\begin{align*}
	\ket{\psi_x(t)} & = \sum_{x'\in \PC}\sum_{\sigma \in \paths}
    \CE(e_t) \dotsm \CE(e_1) \ket{\psi_{x'}(0)}.
\end{align*}

The situation of main interest for us is when the underlying graph of the walk
is actually a Cayley graph $\Cay{G,S}$.  In such a case, the transition matrices
depend only on the generators associated to the edges and, as before, we let
$\CE(xh,x) = A_h$.  Notice that a path $\sigma$ is in $\paths$ if and only if
there exist $h_1,\dots,h_t \in S$ such that $x'h_1\dotsm h_t = x$; therefore we
obtain the expression
\begin{equation}\label{eq:path-sum-g}
	\ket{\psi_x(t)} = \sum_{x' \in \PC}
    	\sum_{h_1,\dots,h_t \in S}
    	\delta(x^{-1} x' h_1 \dotsm h_t)
    	\CA(h_1^{-1},\dots,h_t^{-1})
    	\ket{\psi_{x'}(0)},
\end{equation}
where $\delta(x) = 1$ if $x = \epsilon$ and $0$ otherwise, $\epsilon$ being the
identity element of $G$, and $\CA(h_1,h_2,\dots,h_t) = A_{h_t} A_{h_{t-1}}
\dotsm A_{h_1}$.

In the following section, we will present an application of the
path-sum method in the specific case of the Weyl QW in $3+1$
dimensions deriving the solution in position space with the aid of a
binary encoding of paths, an approach which has already proven its
efficacy in the case of the Dirac QW in $1+1$ dimensions
\cite{DAriano:2014ad} and of the
Weyl QW in $2+1$ dimensions \cite{DAriano:2015aa}.

\section{Weyl quantum walk in $3+1$ dimensions} \label{sec:weyl-qw}

In the following we will use the vector notation to denotes points $\bx\in G=\Z^3$.
The Weyl \QW in $3+1$ dimensions, derived by D'Ariano and
Perinotti~\cite{DAriano:2014ae}, is a \QW on the \BCC lattice $\Cay{G,S}$, where
the vertex set can be chosen to be $G = 2\Z^3 \cup (2\Z^3 + (1,1,1))$ and the
generators can be represented by the following vectors (and their inverses):
\begin{equation*}
	\bh_{1} = \mvec{ 1\\ 1\\ 1}, \quad
	\bh_{2} = \mvec{ 1\\-1\\-1}, \quad
	\bh_{3} = \mvec{-1\\ 1\\-1}, \quad
	\bh_{4} = \mvec{-1\\-1\\ 1}.
\end{equation*}
The walk unitary operator is written as $W = \sum_{\bh \in S} T_{\bh} \otimes
A_{\bh}$, acting on $\ell^2(G) \otimes \C^2$, and the transition matrices are
\begin{equation*}
\begin{aligned}
    A_{\bh_1} & = \zeta^*
    \begin{pmatrix}
        1 & 0 \\
        1 & 0
    \end{pmatrix}, & \quad
    A_{\bh_{-1}} & = \zeta
    \begin{pmatrix}
        0 & -1 \\
        0 &  1
    \end{pmatrix}, \\
    A_{\bh_2} & = \zeta^*
    \begin{pmatrix}
        0 & 1 \\
        0 & 1
    \end{pmatrix}, & \quad
    A_{\bh_{-2}} & = \zeta
    \begin{pmatrix}
         1 & 0 \\
        -1 & 0
    \end{pmatrix}, \\
    A_{\bh_3} & = \zeta^*
    \begin{pmatrix}
        0 & -1 \\
        0 &  1
    \end{pmatrix}, & \quad
    A_{\bh_{-3}} & = \zeta
    \begin{pmatrix}
        1 & 0 \\
        1 & 0
    \end{pmatrix}, \\
    A_{\bh_4} & = \zeta^*
    \begin{pmatrix}
         1 & 0 \\
        -1 & 0
    \end{pmatrix}, & \quad
    A_{\bh_{-4}} & = \zeta
    \begin{pmatrix}
        0 & 1 \\
        0 & 1
    \end{pmatrix},
\end{aligned}
\end{equation*}
with $\zeta = \frac{1 \pm \im}{4}$ and $\bh_{-l} = -\bh_l$. The two possible
choices for the coefficient $\zeta$ correspond to the two inequivalent \QW
solutions existing on the \BCC lattice~\cite{DAriano:2014ae}.

The intent of this paper is to provide an explicit expression for the propagator
given in terms of sums over paths. To this end, it turns out to be very
effective to adopt a binary description of paths, which amounts to find a
three-bits binary encoding $b_1b_2b_3$ for the generators. A suitable choice
that simplifies the evaluation of the contribution to the transition amplitude
of each path is given by
\begin{equation}\label{eq:enc}
\begin{aligned}
    & \bh_1 \colon 011, &
    & \bh_2 \colon 110, &
    & \bh_3 \colon 101, &
    & \bh_4 \colon 000, \\
    & \bh_{-1} \colon 100, &
    & \bh_{-2} \colon 001, &
    & \bh_{-3} \colon 010, &
    & \bh_{-4} \colon 111.
\end{aligned}
\end{equation}
This choice is tantamount to the requirement
\begin{align*}
	\tilde A_{b_1b_2b_3} =
    (\pm\im)^{b_1 \oplus b_2 \oplus b_3} B_{b_1b_2}, \qquad
  \tilde A_{b_1b_2b_3} \deq (\zeta^*)^{-1} A_{b_1b_2b_3},
\end{align*}
where $\oplus$ denotes the sum modulo $2$ and $B_{b_1b_2}$ are the matrices
\begin{equation*}
\begin{aligned}
	B_{00} & =
	\begin{pmatrix}
	 1 & 0 \\
	-1 & 0
	\end{pmatrix}, & \quad
	B_{10} & =
	\begin{pmatrix}
	0 & -1 \\
	0 &  1
	\end{pmatrix}, \\
	B_{01} & =
	\begin{pmatrix}
	1 & 0 \\
	1 & 0
	\end{pmatrix}, & \quad
	B_{11} & =
	\begin{pmatrix}
	0 & 1 \\
	0 & 1
	\end{pmatrix},
\end{aligned}
\end{equation*}
satisfying the product rule given by
\begin{equation}\label{eq:mat-2d-prod}
    B_{ab} B_{cd} = (-1)^{(c \oplus a)\cdot(d \oplus b)} B_{cb},
\end{equation}
where the notation $a\cdot b$ denotes the binary product of the two bits $a$ and
$b$; throughout this paper, all the bit operations are also extended
element-wise to binary strings.
One can notice from \cref{eq:mat-2d-prod} that the matrices
$\tilde{A}_{b_1b_2b_3}$ generate, up to phases, a finite semigroup.
The general product of $t$ transition matrices is then given by
\begin{align}
  \tilde{\CA}\wtre
    & \deq \tilde{A}_{\ws1_t\ws2_t\ws3_t} \cdots
           \tilde{A}_{\ws1_1\ws2_1\ws3_1}\nonumber \\
    & = (-1)^{\iota((\ws1 \oplus S\ws1) \cdot \ws2)}
        (\pm\im)^{\iota(\ws1 \oplus \ws2 \oplus \ws3)}
        B_{\ws1_{1}\ws2_{t}}, \label{eq:mat-prod}
\end{align}
where $\ws{j} \in \Bin t$ is the string made of the $j$-th bits of the encoding
\cref{eq:enc}, $\Bin t$ being the set of all binary strings of length
$t$. Here $w_k$ denotes the $k$-th bit of the string $w$.  Finally we
introduced the function $\iota(w) \deq \sum_{k=1}^{\len{w}} w_k$
counting the number of $1$-bits present in the string $w$ and the left
circular shift $S$ defined by $(Sw)_i = w_{(i \bmod t) + 1}$, for all
$i=1,\dots,t$.

At this point, we need a characterization of the lattice paths in
terms of their binary description.  As shown in
\cref{app:set-bit-counts}, the number of $1$-bits in each one of the
encoding strings $\ws{j}\in\Bin t$ is fixed by the starting point
$\bx'$ and by the ending point $\bx$ via the equations
\begin{equation*}
\begin{cases}
    \iota(\ws1) = \frac{t - (x^3-x'^3)}{2}, \\
    \iota(\ws2) = \frac{t + (x^1-x'^1)}{2}, \\
    \iota(\ws3) = \frac{t + (x^2-x'^2)}{2}.
\end{cases}
\end{equation*}
It is then convenient to define a special notation for the set of tuples of
strings with a fixed number of $1$-bits, which are indeed in bijective correspondence with paths connecting $\bx'$ to $\bx$ in $t$ steps:
\begin{align*}
    \BT t n {K_1,\dots,K_n} & \deq
        \Set{\wn \in \Bin t^n |
             \iota(w^{(i)}) = K_i, \;
             \forall \sp1 i = 1,\dots,n} \\
    & = \BS t{K_1} \times \dots \times \BS t{K_n},
\end{align*}
assuming for short $\BS tK \equiv \BT t1K$.
Thus, the path-sum of \cref{eq:path-sum-g} can be rewritten as
\begin{gather}\label{eq:path-sum-bin}
	\ket{\psi_\bx(t)}
    = {\zeta^*}^{t} \sum_{\bx' \in \PC[\bx]}\;
        {\sum_{\vec{w}\in\BT t3{K_1,K_2,K_3} }}
        \tilde{\CA}(\vec{w})
        \ket{\psi_{\bx'}(0)},
\end{gather}
where $K_1=\frac{t+(x^3-x'^3)}{2},K_2=
             \frac{t-(x^1-x'^1)}{2},K_3=\frac{t-(x^2-x'^2)}{2}$
(notice that the signs here are consistent with \cref{eq:path-sum-g} where
the matrix associated to the step $h$ is exactly $A_{h^{-1}}$).  From
\cref{eq:mat-prod}, we see that the resulting matrix depends only on the first
bit of $\ws1$ and the last bit of $\ws2$.  Therefore, defining
$\BT{t,ab}3{K_1,K_2,K_3} \subset \BT t3{K_1,K_2,K_3}$ as the subset of triads
$(\ws1,\ws2,\ws3)$ with $\ws1_1 = a$ and $\ws2_t = b$, we can rewrite
\cref{eq:path-sum-bin} as:
\begin{gather}
	\ket{\psi_\bx(t)}
     = {\zeta^*}^{t} \sum_{\bx' \in \PC[\bx]} \sum_{a,b=0,1}
        c_{ab} B_{ab} \ket{\psi_{\bx'}(0)}, \label{eq:path-sum-fixed} \\
    \shortintertext{where the coefficients $c_{ab}$ are defined by}
    c_{ab}(K_1,K_2,K_3) = \sum_{\vec{w}\in\BT{t,ab}3{K_1,K_2,K_3}}
	    (-1)^{\iota((\ws1\oplus S\ws1)\cdot\ws2)}
		    (\pm\im)^{\iota(\ws1 \oplus \ws2 \oplus \ws3)}. \label{eq:cof}
\end{gather}
%
%provided that the sum is computed over the set $\BT{t,ab}3{K_1,K_2,K_3}$.

We seek now an explicit expression for the coefficients $c_{ab}$.  As a first
step let us consider the sum $\sum_{w} (\pm\im)^{\iota(v \oplus w)}$ alone with
$\iota(w) = H$ for a given $H$ and for some fixed $v \in \BS tK$.  In order to
compute this sum we have to classify the strings lying in the set $v \oplus \BS
tH$, according to the number of ones.  It is convenient to define a canonical
form for strings as follows.
\begin{definition}\label{def:cform1}
We say that a binary string $w$ is in \emph{canonical form} if the following condition holds:
\begin{align*}
	w_i \geq w_{i+1} \quad \forall\sp1 i = 1,2,\dots,\len w - 1.
\end{align*}
\end{definition}

The computation of the coefficients $c_{ab}$ in Eq.~\eqref{eq:cof} is
based on the next three results whose proofs are presented in Appendix
\ref{app:proofs}.
\begin{lemma}\label{lem:split}
Given $t,K,H \in \N$, with $K,H\leq t$, let $v \in \BS tK$ be in canonical form, then we have
\begin{gather}
	v \oplus \BS tH = \bigsqcup_{n\in I} \BWn, \\
	\BWn\deq
	\begin{cases}
		\BS K{K-H+n} \BS{t-K}n, & \text{if $K \geq H$}, \\
		\BS K{n} \BS{t-K}{H-K+n}, & \text{otherwise},
	\end{cases}
\end{gather}
with $I = \{0,1,\dots,\min\{K,H,t-K,t-H\}\}$.
Defining $r = \theta(K-H)$ and $\bar r = 1-r$, the size of each subset is given by
\begin{gather}
	\card{\BWn} = D(t, \bar r t + (-1)^{\bar r} H, rt + (-1)^r K, n), \\
	D(t,p,m,n) =
	\begin{cases}
		\binom{m}{n} \binom{t-m}{p-n}, &
			\text{if $0 \leq n \leq m \leq t$ and $n\leq p$,} \\
		0, & \text{otherwise},
	\end{cases} \nonumber
\end{gather}
and for all $w \in \BWn$ the set-bit count is given by $\iota(w) = \abs{K-H} + 2n$.
\end{lemma}

\begin{corollary}\label{cor:split}
Given $t,K,H \in \N$, with $K,H\leq t$, for all $v \in \BS tK$ and for any $\pi_v$ bitwise permutation of $\Bin t$ such that $\pi_v(v)$ is in canonical form, the following decomposition holds:
\begin{gather*}
	v \oplus \BS tH = \bigsqcup_{n\in I} \pi_v^{-1}(\BWn) \\
\shortintertext{with $I = \{0,1,\dots,\min\{K,H,t-K,t-H\}\}$ and moreover}
\begin{aligned}
	\card{\pi_v^{-1}(\BWn)} & = \card{\BWn}, \\
	\iota(\pi_v^{-1}(\BWn)) & = \iota(\BWn).
\end{aligned}
\end{gather*}
\end{corollary}

\begin{proposition} \label{prp:count-w3}
Given $t,K,H \in \N$, with $K,H\leq t$, for all $v \in \BS tK$ we have that
\begin{gather}
	\sum_{w \in \BS tH} (\pm\im)^{\iota(v \oplus w)} = f(K,H), \label{eq:sum-3} \\
\shortintertext{where}
	f(K,H) \deq
	        (\pm\im)^{\abs{K - H}}
	        \sum_{n\in I} (-1)^n
	        D(t, \bar r t + (-1)^{\bar r} H, rt + (-1)^r K, n), \nonumber
\end{gather}
with $I = \{0,1,\dots,\min\{K,H,t-K,t-H\}\}$.
\end{proposition}

The expression of the coefficients $c_{ab}$ of \cref{eq:cof} can now
be written as
\begin{equation}\label{eq:cof-sum-12}
	c_{ab}(K_1,K_2,K_3) = \sum_{\ws1,\ws2}
		(-1)^{\iota((\ws1\oplus S\ws1)\cdot\ws2)}
    f\pp{\iota\pp{\ws1 \oplus \ws2},K_3},
\end{equation}
where, at this point, the sum is performed over the set
$\BT{t,ab}2{K_1,K_2}$.  In order to ease the calculation we can study
separately the values of $\iota((\ws1\oplus S\ws1)\cdot\ws2)$ and
their interplay with $f(K,H)$.  The combinatorics for
$\iota((\ws1\oplus S\ws1)\cdot\ws2)$ is the same as that of
Ref.~\cite{DAriano:2015aa}, and is here reviewed for the convenience
of the reader. The idea is to find a classification of the binary
strings in terms of the values taken by
$\iota((\ws1\oplus S\ws1)\cdot\ws2)$.  We first focus on the
classification induced by $\ws1\oplus S\ws1$ and then, since the only
contributions to the result come from the $1$-bits of both
$\ws1\oplus S\ws1$ and $\ws2$, we use $\ws2$ to select a given number
of $1$-bits from $\ws1\oplus S\ws1$.

The combinatorics induced by $\ws1\oplus S\ws1$ is provided by the following
lemma (proved in Appendix \ref{app:proofs}).

\begin{lemma}\label{lem:c-interf}
Let $v \in \BS tK$. Then $v \oplus Sv \in \pi_v^{-1}\pp{\BS Kn \BS{t-K}n}$,
for some $n \in \{0,1,\dots,\min\{K,t-K\}\}$.
\end{lemma}

The above classification is not sufficient in our case, since the summation of
\cref{eq:cof-sum-12} requires the strings $\ws1$ ad $\ws2$ to have,
respectively, the first and the last bit fixed.  This means that we have to
refine the counting taking into account such constraint: we denote $\BSTn$ the
set of strings $v \in \BS tK$ such that $v \oplus Sv \in \pi_v^{-1}\pp{\BS Kn
\BS{t-K}n}$ with $v_1 = a$ and $v_t = a'$.  The following lemma (also proved in
Appendix \ref{app:proofs}) concludes the counting for strings with fixed
endpoints.

\begin{lemma}\label{lem:c-interf-fixed}
The number $u_{aa'}(n)$ of binary strings $v\in\BSTn$ is given by
\begin{align}
	u_{aa'}(n) \coloneqq \card{\BSTn} = C_{K,n+aa'} C_{t-K,n+\bar a \bar a'},
\end{align}
with $n\tp{min}(K) \leq n \leq n\tp{max}(K)$ and $n\tp{min}(K)\deq\min\{1,K,t-K\}$,
\begin{gather}
	n\tp{max}(K)\deq
	\begin{cases}
		\min\{K-aa',t-K-1+aa'\}, & \text{if $1 < K < t-1$}, \\
		1, & \text{if $K=1$ or $K=t-1$}, \\
		0, & \text{otherwise.}
	\end{cases}
\end{gather}
\end{lemma}

Finally we can focus on the joint classification given by the values of
$\iota((v \oplus Sv) \cdot w)$ and of $\iota(v \oplus w)$. The final expression
of the coefficients $c_{ab}$ is summarized by the following proposition, whose
proof is reported in \cref{app:proofs}.

\begin{proposition}\label{prp:count-w12}
For $t\geq 2$ the coefficients of \cref{eq:cof-sum-12} take the form
\begin{gather}
	c_{ab}(K_1,K_2,K_3) =
	\sum_{a'=0,1} \sum_{n=n\tp{min}(K_1)}^{n\tp{max}(K_1)}
	\sum_{J=b}^{K_2} u_{aa'}(n)
	w_{aa'b}^{(0)}(n,J)w_{aa'b}^{(1)}(n,J) f(\tau_J,K_3), \label{eq:cof-final}
\end{gather}
where, defining $r = \theta(K_1-K_2)$, we have that
$\tau_J = \abs{K_1 - K_2} + 2J$, and
\begin{align}
	& w_{aa'b}^{(s)}(n,J) =
		\sum_{k=0}^{n-\gamma_{aa'}^{(s)}} (-1)^{k+\gamma_{aa'}^{(s)}b}
		D(\eta_{a'}^{(s)}, \kappa_{a'b}^{(s)}(J),
			n-\gamma_{aa'}^{(s)}, k), \label{eq:ws} \\
	& \kappa_{a'b}^{(s)}(J) =
	(r\oplus\bar s) K_2 +
	(-1)^{\bar s}(\bar r K_1-J)-(\bar s \oplus a')b, \nonumber \\
	& \eta_{a'}^{(s)} = \bar s (t-1) + (-1)^{\bar s}(K_1-a'), \quad
	\gamma_{aa'}^{(s)} = (s\oplus a)(\bar s \oplus a'). \nonumber
\end{align}
\end{proposition}

As a concluding remark of this section we point out that thanks to the binary
encoding of paths, the counting is largely simplified providing a powerful tool
working in the general case.

\section{Conclusions}

In this paper we provided an explicit analytical solution in position space of
the propagator for the Weyl \QW in $3+1$ dimensions.
%As we will see in the details, a QW on the Hilbert space
%$\mathcal{H}_X\otimes \mathcal{H}_C$ ($\mathcal{H}_C$ and
%$\mathcal{H}_X$ representing the coin and spatial Hilbert space
%respectively) can always be written in the form
%$\sum_j T_j\otimes A_j$. Here the index $j$ spans a finite set $J$
%made of the possible jumping ``directions'', with $T_j$ the
%translation operator in that direction.  When the jumping occurs in
%direction $j$, the action on the coin space is given by the
%\emph{transition matrix} $A_j$.
The method used here to analytically compute the Weyl QW path sum is grounded on:

\begin{enumerate}
  \item A combinatorial analysis for the characterization of the admissible paths connecting two fixed sites of the graph in a given number of steps;
  \item The semigroup structure of the walk transition matrixes.
\end{enumerate}

The semigroup structure allows us to classify the admissible paths (or
binary strings according to the first point) into equivalence classes
according to the global transition matrix of the path.
To aid the calculation, we adopted here an extremely powerful approach consisting in the translation of the
geometric properties of lattice paths into algebraic properties of binary
strings. Such binary description of paths was succesful also in the case of the
Dirac \QW in $1+1$ dimensions \cite{DAriano:2014ad} and of the Weyl \QW in $2+1$
dimensions \cite{DAriano:2015aa}. In all these cases the practical computation
of the propagator relies on the semigroup property of the transition matrices
which allows one to determine all the paths contributing with the same matrix.

As a future perspective we propose the possibility to investigate such semigroup
property of the transition matrices in the general non Abelian case. It would also be interesting
to derive the general hypotheses in the framework of \QW{s} on Cayley graphs for
which the semigroup structure is recovered.

\enlargethispage{20pt}

%\ethics{Insert ethics statement here if applicable.}
%\dataccess{Insert details of how to access any supporting data here.}

\aucontribute{Giacomo Mauro D'Ariano and Paolo Perinotti conceived and designed the study. Nicola Mosco carried out the calculations. Nicola Mosco and Alessandro Tosini drafted the manuscript. All the authors read, edited and approved the manuscript.}

\competing{The author(s) declare that they have no competing interests.}

\funding{This publication  was made possible through the support of a grant  from the John Templeton Foundation, ID \# 60609 ``Quantum Causal Structures''. The opinions expressed in this publication are those of the authors and do not necessarily reflect the views of the John Templeton Foundation.
}

%\ack{Insert acknowledgment text here.}

%\disclaimer{Insert disclaimer text here if applicable.}

%%%%%%%%%% Insert bibliography here %%%%%%%%%%%%%%

%\bibliographystyle{rsta}
\bibliographystyle{unsrt}
\bibliography{bibliography}

%%%%%%%%%%%%%%%%%%%%%%%%%%%%%%%%%%%%%%%%%%%%%%%%%%%%%%%%%%%%%%%%%%%%%%%%%%%%%
%%%                        Proof of main results                          %%%
%%%%%%%%%%%%%%%%%%%%%%%%%%%%%%%%%%%%%%%%%%%%%%%%%%%%%%%%%%%%%%%%%%%%%%%%%%%%%

\appendix

\section{Relation between lattice points and binary strings}
\label{app:set-bit-counts}

The approach we follow here requires to find a binary description of paths: we
provided such description by choosing a suitable binary encoding in
\cref{eq:enc} which also simplifies the evaluation of functions involving
products of matrices giving the general formula of \cref{eq:mat-prod}. Then, we
also need to obtain a criterion telling us which paths have the same endpoints
$\bx'$ and $\bx$; in order to do so, we employ the fact that the \BCC lattice is
Abelian: each path is described by a quadruple $(n_{\pm 1}, n_{\pm 2}, n_{\pm
3}, n_{\pm 4})$, where $n_{\pm l}$ counts the number of steps in direction
$\bh_{\pm l}$ and thus each path must satisfy the system
\begin{equation}\label{eq:path-sys}
\begin{cases}
	\sum_{l} (n_{l} - n_{-l}) h_{l}^{i} =
		x^{i} - x'^{i}, \\
	\sum_{l} (n_{l} + n_{-l}) = t.
\end{cases}
\end{equation}
The binary encoding and the quadruple $(n_{\pm 1}, n_{\pm 2}, n_{\pm 3}, n_{\pm
4})$ are then related by the following constraints:
\begin{equation}\label{eq:iotas}
\begin{cases}
    \iota(\ws1) = n_{-1} + n_{2} + n_{3}  + n_{-4}, \\
    \iota(\ws2) = n_{1}  + n_{2} + n_{-3} + n_{-4}, \\
    \iota(\ws3) = n_{1}  + n_{3} + n_{-2} + n_{-4},
\end{cases}
\end{equation}
where $\iota(w) \deq \sum_{k=1}^{\len{w}} w_k$ denotes the set-bit count of the
string $w$.  The relation we seek can be easily obtained by defining the number
of steps in the positive- and negative-coordinate direction
\begin{align*}
  x_+^i & \deq \sum_l \sparen*{\theta(h_l^i) n_l + \theta(h_{-l}^i) n_{-l}}, \\
  x_-^i & \deq \sum_l \sparen*{\theta(-h_l^i) n_l + \theta(-h_{-l}^i) n_{-l}},
\end{align*}
$\theta$ being the Heaviside step function; then, for each $i$ it holds that
\begin{equation}\label{eq:xpm}
\begin{cases}
    x_+^i + x_-^i = t, \\
    x_+^i - x_-^i = x^i - x'^i.
\end{cases}
\end{equation}
Therefore, the set-bit counts of \cref{eq:iotas} are fixed
by the coordinates through the equations
\begin{equation}
\begin{cases}
    \iota(\ws1) = x_-^3 = \frac{t - (x^3-x'^3)}{2}, \\
    \iota(\ws2) = x_+^1 = \frac{t + (x^1-x'^1)}{2}, \\
    \iota(\ws3) = x_+^2 = \frac{t + (x^2-x'^2)}{2}.
\end{cases}
\end{equation}

\section{Proofs of the results} \label{app:proofs}

\begin{proof}[Proof of \cref{lem:split}]
Suppose for definiteness that $K \geq H$.
Given a string $w \in \BS tH$ we can define the two substrings $(L_Kw)_i = w_i$,
for $i = 1,\dots,K$, and $(R_Kw)_i = w_{K+i}$, for $i = 1,\dots,t-K$: we have
that $L_Kw \in \BS K{H-n}$, $R_Kw \in \BS {t-K}n$ and $w \in \BS K{H-n} \BS
{t-K}n$. Here for $A$ and $B$ sets of strings, the set $AB$ is formed by
concatenating the elements of $A$ and $B$.
Moreover if $n \neq m$, then $\BS Kn \cap \BS Km = \emptyset$, which entails
that the products $\BS K{H-n} \BS {t-K}n$ form a partition of $\BS tH$, with $n
\in I = \{0,1,\dots,\min\{H,t-K\}\}$:
\begin{gather}\label{eq:split-k}
	\BS tH = \bigsqcup_{n\in I} \BS K{H-n} \BS {t-K}n.
\end{gather}
Thus the set $v \oplus \BS tH$ decomposes as
\begin{align*}
	v \oplus \BS tH & = \bigsqcup_{n\in I} v\oplus[\BS K{H-n} \BS {t-K}n] \\
					& = \bigsqcup_{n\in I} [L_Kv \oplus \BS K{H-n}]
							[R_Kv \oplus \BS {t-K}n] \\
					& = \bigsqcup_{n\in I} \BS K{K-H+n} \BS {t-K}n \\
					& = \bigsqcup_{n\in I} \BWn,
\end{align*}
where in the third equation we have made use of the fact that $v$ is in canonical form.
The set-bit count of $w \in \BWn$ can be straightforwardly computed:
\begin{align*}
	\iota(w) = \iota(L_Kw) + \iota(R_Kw) = K - H + 2n.
\end{align*}
Finally, one can easily obtain the size of each factor:
\begin{align*}
	\card{\BWn} & = \card{\BS K{K-H+n} \BS {t-K}n} \\
				& = \card{\BS K{K-H+n}} \cdot \card{\BS {t-K}n} \\
				& = \binom{K}{H-n}\binom{t-K}{n}.
\end{align*}
The case $K < H$ follows from a similar reasoning and we have the sought result.
\end{proof}

\begin{proof}[Proof of \cref{cor:split}]
Let $v \in \BS tK$ and let $\pi_v$ be a bitwise permutation such that $\pi_v(v)$ is canonical, then we have that
\begin{align*}
	\pi_v(v \oplus \BS tH)
		& = \pi_v(v) \oplus \pi_v(\BS tH) \\
		& = \pi_v(v) \oplus \BS tH.
\end{align*}
Therefore, using the decomposition of \cref{lem:split}, we can write
\begin{align*}
	v \oplus \BS tH
		& = \pi_v^{-1}\pp{\bigsqcup_{n\in I} \BWn} \\
		& = \bigsqcup_{n\in I} \pi_v^{-1}(\BWn). \qedhere
\end{align*}
\end{proof}

\begin{proof}[Proof of \cref{prp:count-w3}]
Suppose that $K \geq H$ and let $v \in \BS tK$, $w \in \BS tH$.
From \cref{cor:split} we know that $v \oplus w \in \pi_v^{-1}(\BWn)$, for some
$n \in I = \{0, 1, \dots, \min\{H,t-K\}\}$, and $\iota(v \oplus w) = K - H +
2n$.
Therefore we can compute the sum of \cref{eq:sum-3}, obtaining finally the
result:
\begin{align*}
	\sum_{w \in \BS tH} (\pm\im)^{\iota(v \oplus w)}
	& = \sum_{n=0}^{\min\{H,t-K\}}
		\sum_{w \in \pi_v^{-1}(\BWn)} (\pm\im)^{\iota(w)} \\
	& = \sum_{n=0}^{\min\{H,t-K\}} \card{\BWn} \, (\pm\im)^{K-H+2n} \\
	& = (\pm\im)^{K-H} \sum_{n=0}^{\min\{H,t-K\}} (-1)^n
		\binom{K}{H-n}\binom{t-K}{n}. \qedhere
\end{align*}
\end{proof}

\begin{proof}[Proof of \cref{lem:c-interf}]
Let $v \in \BS tK$. From \cref{lem:split,cor:split} we know that $v \oplus Sv \in \BS t{2n}$, for some $n \in\{0,1,\dots,\min\{K,t-K\}\}$. We can split the string $\pi_v(v\oplus Sv) \eqqcolon c \oplus \pi_vSv$ as
\begin{align*}
	L_K(c\oplus\pi_vSv) & = L_Kc \oplus L_K\pi_vSv = \overline{L_K\pi_vSv}, \\
	R_K(c\oplus\pi_vSv) & = R_Kc \oplus R_K\pi_vSv = R_K\pi_vSv.
\end{align*}
Given that
\begin{gather*}
	\iota(L_K\pi_vSv) + \iota(R_K\pi_vSv) = \iota(v) = K, \\
\intertext{the set-bit count of $c\oplus\pi_vSv$ then reads}
\begin{aligned}
	\iota(c\oplus\pi_vSv)
	& = \iota(\overline{L_K\pi_vSv}) + \iota(R_K\pi_vSv) \\
	& = 2K - 2\iota(L_K\pi_vSv).
\end{aligned}
\end{gather*}
Therefore, since $\iota(c\oplus\pi_vSv) = 2n$, we have precisely
\begin{gather*}
	\iota(\overline{L_K\pi_vSv}) = \iota(R_K\pi_vSv) = n
\end{gather*}
and the result follows.
\end{proof}

\begin{proof}[Proof of \cref{lem:c-interf-fixed}]
A string in $\BSTn$ can be constructed by first arranging the $K$ $1$-bits in $n
+ aa'$ slots and then by arranging the $t-K$ $0$-bits in $n + \bar a \bar a'$
slots: the problem is then the same as counting the number of compositions of an
integer~\cite{feller:2008a}.  We denote here as $C_{K,n}$ the number of
$n$-compositions of an integer $K$:
\begin{align*}
	C_{K,n} =
	\begin{cases}
		\binom{K-1}{n-1}, & \text{if $K\geq n>0$,} \\
		1, & \text{if $K=n=0$,} \\
		0, & \text{otherwise.}
	\end{cases}
\end{align*}
Since the $n+aa'$ slots can be filled independently from the $n + \bar a \bar a'$ ones, the cardinality of $\BSTn$ is simply given by the product
\begin{equation*}
  u_{aa'}(n) \deq \card{\BSTn} =
    C_{K,n+aa'} C_{t-K,n + \bar a \bar a'}. \qedhere
\end{equation*}
\end{proof}

\begin{proof}[Proof of \cref{prp:count-w12}]
Let us consider the case where $K_1 \geq K_2$ as the other one is a simple
variation of the following construction.  First of all, recall the expression
for the coefficients~\eqref{eq:cof-sum-12}, implementing the result of
\cref{prp:count-w3}:
\begin{align*}
	c_{ab}
    & = \sum_{v,v'} (-1)^{\iota((v\oplus Sv)\cdot v')}
        f(\iota(v\oplus v'), K_3) \\
    & = \sum_{a'=0,1} \sum_{n=n\tp{min}(K_1)}^{n\tp{max}(K_1)}
        \sum_{v_n} \sum_{w} (-1)^{\iota((av_na'\oplus v_na'a)\cdot wb)}
            f(\iota(av_na'\oplus wb), K_3),
\end{align*}
where $av_na' \in \BST{aa'}{t}{K_1}{n}$ and $w\in\BS{t-1}{K_2-b}$.
For ease of discussion, we consider from now on the case where $a=a'=0$, as
the others can be treated in a very similar way.

We already know that the sum modulo $2$ of a pair of strings $0v_n0$ and $wb$
can be parametrized as $\iota(0v_n0\oplus wb) = |K_1-K_2| + 2J$ for some $J$;
yet we need to find a finer classification taking into account the values of
$\iota((0v_n0\oplus v_n00)\cdot wb)$, for each fixed $v_n$. To such end, we
consider the partition
\begin{align}
  \BS{t-1}{K_2-b} & = \bigsqcup_{J=b}^{\min\{K_2,t-K_1-1\}}
  	\bigsqcup_{k=0}^n \bigsqcup_{k'=0}^n \BZn1 \BZ b0{J,k'}, \nonumber \\
	\BZn1 & =
		\BS nk \BS{K_1-n}{K_2-J-k}, \label{eq:bz1} \\
  \BZ b0{J,k'} & =
		\BS n{k'} \BS{t-K_1-n-1}{J-k'-b}; \label{eq:bz0}
\end{align}
this particular construction corresponds to choose as reference strings
\begin{align*}
  p_L = \underbrace{11\cdots 1}_{n}
    \underbrace{00\cdots 0}_{K_1-n}, \qquad
  p_R = \underbrace{11\cdots 1}_{n}
    \underbrace{00\cdots 0}_{t-K_1-n-1},
\end{align*}
where the text under braces denotes the length of the substring.

Since $\BS t{K_2}$ can be constructed by concatenating strings picked from the
sets $\BZ b1{J,k}$ and $\BZ b0{J,k'}$, we can study separately the binary
product and binary sum on them.  Multiplying $p_L$ and $p_R$ with $\BZ b1{J,k}$
and $\BZ b0{J,k'}$ respectively, we obtain
\begin{gather*}
    p_L \cdot \BZ b1{J,k} = \BS nk \underbrace{00\cdots 0}_{K_1-n}, \\
    p_R \cdot \BZ b0{J,k'} = \BS n{k'} \underbrace{00\cdots 0}_{t-K_1-n-1}, \\
    \iota\pp{(p_L p_R) \cdot w} = k + k',
\end{gather*}
for all $w\in\BZ b1{J,k} \BZ b0{J,k'}$; whereas, for the binary sum we get
\begin{gather*}
    \iota\Big(\underbrace{11\cdots 1}_{K_1}\oplus\BZ b1{J,k}\Big)
      = K_1 - K_2 + J, \qquad
    \iota\pp{\BZ b0{J,k'}}
      = J - b.
\end{gather*}
In conclusion, the total number of ones for the sum modulo $2$ turns out to be
\begin{equation*}
	\iota\pp{[\pi_{t-2}(v_n)00] \oplus \sparen*{\BZ b1{J,k}\BZ b0{J,k'}b}} =
    K_1 - K_2 + 2J,
\end{equation*}
which corresponds to the parameter $\tau_J$ of the statement of the proposition.
Furthermore, since $f(\iota(0v_n0 \oplus wb),K_3)$ does not depend on $k$ nor
$k'$, we can sum separately over the strings in \cref{eq:bz0,eq:bz1} obtaining
the function $w_{00b}^{(s)}$ of \cref{eq:ws}:
\begin{equation*}
  w_{00b}^{(s)}(n,J) = \sum_{k=0}^n (-1)^k \card*{\BZ bs{J,k}}. \qedhere
\end{equation*}
\end{proof}

\end{document}